\documentclass[12pt]{article}

\usepackage{epsfig}

\usepackage{bbm}
\usepackage{amssymb,amsmath}

\newcommand{\be}{\begin{equation}}
\newcommand{\ee}{\end{equation}}
\newcommand{\ba}{\begin{eqnarray}}
\newcommand{\ea}{\end{eqnarray}}
\newcommand{\no}{\nonumber\\}

\newcommand{\mnu}{\mathcal{M}_\nu}

\newcommand{\dsol}{\Delta m^2_\odot}
\newcommand{\datm}{\Delta m^2_\mathrm{atm}}
\newcommand{\zz}{\mathbbm{Z}_2}
\newcommand{\zzz}{\mathbbm{Z}_3}

\newcommand{\ut}{\underline{3}}
\newcommand{\uo}{\underline{1}}
\newcommand{\utwo}{\underline{2}}

\begin{document}

\title{\normalsize \hfill UWThPh-2008-13 \\[8mm]
\LARGE A model for trimaximal lepton mixing}

\author{
W.~Grimus$^{(1)}$\thanks{E-mail: walter.grimus@univie.ac.at}
\ and
L.~Lavoura$^{(2)}$\thanks{E-mail: balio@cftp.ist.utl.pt}
\\*[3mm]
$^{(1)}$ \small
University of Vienna, Faculty of Physics \\
\small
Boltzmanngasse 5, A--1090 Vienna, Austria
\\*[2mm]
$^{(2)}$ \small
Universidade T\'ecnica de Lisboa
and Centro de F\'\i sica Te\'orica de Part\'\i culas \\
\small
Instituto Superior T\'ecnico, 1049-001 Lisboa, Portugal
}

%%%%
\date{9 September 2008}

\maketitle

\begin{abstract}
We consider trimaximal lepton mixing,
defined by $\left| U_{\alpha 2} \right|^2 = 1/3 \
\forall \alpha = e, \mu, \tau$.
This corresponds to a two-parameter lepton mixing matrix $U$.
We present a model for the lepton sector
in which trimaximal mixing is enforced 
by softly broken discrete symmetries; 
one version of the model is based on the group $\Delta (27)$.
A salient feature of our model is that no vacuum alignment is required.
\end{abstract}

\newpage

\section{Introduction}

It is now experimentally firmly established
that the neutrinos are massive
and that leptons of different families
mix in the charged weak 
interaction---see~\cite{reviews} for reviews and~\cite{fogli,tortola}
for recent fits. 
Lepton mixing is given by a $3 \times 3$ unitary matrix
$\left( U_{\alpha j} \right)$,
with $\alpha = e, \mu, \tau$ corresponding to the lepton flavours
and $j = 1, 2, 3$ corresponding to the neutrino mass eigenstates.
According to the most recent
three-flavour neutrino oscillation update~\cite{tortola}, 
at the $3\, \sigma$ level,
the results for neutrino mixing are
\be
\left| U_{e3} \right|^2 \le 0.056,
\label{urtye}
\ee
\be
0.25 \le \sin^2{\theta_\odot} \le 0.37,
\ee
\be
0.36 \le \sin^2{\theta_\mathrm{atm}} \le 0.67,
\label{njdft}
\ee
where
\ba\label{sinsol}
\sin^2{\theta_\odot} &\equiv&
\frac{\left| U_{e2} \right|^2}
{1 - \left| U_{e3} \right|^2},
\\\label{atmsol}
\sin^2{\theta_\mathrm{atm}} &\equiv&
\frac{\left| U_{\mu 3} \right|^2}
{1 - \left| U_{e3} \right|^2}.
\ea
Moreover,
at the $3 \, \sigma$ level
the neutrino mass differences satisfy~\cite{tortola}
\be
7.05 \times 10^{-5}\, \mathrm{eV}^2
\le \dsol \equiv m_2^2 - m_1^2 \le
8.34 \times 10^{-5}\, \mathrm{eV}^2,
\ee
\be
2.07 \times 10^{-3}\, \mathrm{eV}^2
\le \datm \equiv \left| m_3^2 - m_1^2 \right| \le
2.75 \times 10^{-3}\, \mathrm{eV}^2,
\label{ndunf}
\ee
the sign of $m_3^2 - m_1^2$ being unknown.

The bounds~(\ref{urtye})--(\ref{njdft}) suggest that
the lepton mixing might be tri-bimaximal,
\textit{i.e.}~that the lepton mixing matrix might be,
apart from unphysical rephasings and from possible Majorana phases,
\be
U = U_\mathrm{HPS} \equiv \left( \begin{array}{rrr}
2/\sqrt{6} & 1/\sqrt{3} & 0 \\ 
-1/\sqrt{6} & 1/\sqrt{3} & -1/\sqrt{2} \\ 
-1/\sqrt{6} & 1/\sqrt{3} & 1/\sqrt{2}
\end{array} \right).
\label{kgure}
\ee
The phenomenological hypothesis $U = U_\mathrm{HPS}$ has
been put forward by Harrison,
Perkins and Scott (HPS)~\cite{HPS}.
However,
it turns out that,
at the model-building level,
it is quite awkward to enforce $U = U_\mathrm{HPS}$ through symmetries.
In particular,
all existing models for $U = U_\mathrm{HPS}$
involve some form of vacuum alignment,
\textit{i.e.}~two subsectors of the scalar sector
having vacuum expectation values (VEVs)
aligned in different directions;
some papers where this problem has been discussed
are found in~\cite{aligned},
%%%%
for more papers see the vast bibliography of~\cite{merlo}.\footnote{An
alternative approach consists in using extra dimensions for model
building, see for instance~\cite{extradim}.} 

In this paper we
adopt the milder hypothesis
that lepton mixing is trimaximal,
\textit{i.e.}~that
\be
\left| U_{e2} \right|^2
= \left| U_{\mu 2} \right|^2
= \left| U_{\tau 2} \right|^2
= \frac{1}{3}.
\ee
Trimaximal mixing relaxes some of the HPS assumptions~\cite{HPS},
since it allows for a nonzero $U_{e3}$
as well as for $\sin^2{\theta_\mathrm{atm}} \neq 1/2$. 
Our main purpose in this paper is to show that trimaximal lepton mixing
may be enforced through a simple model which involves no vacuum alignment.

In section~\ref{phenomenology} we make
a brief phenomenological study of trimaximal mixing,
proceeding in section~\ref{model} to present
the simplest version of our model.
In section~\ref{extensions}
we consider some variations on the model.
Our conclusions are found in section~\ref{conclusions}.

\section{Trimaximal mixing}
\label{phenomenology}

It follows from the trimaximal-mixing assumption
$\left| U_{e2} \right|^2 = 1/3$ and equation~(\ref{sinsol}) that
\be\label{solmix}
\sin^2{\theta_\odot}
= \frac{1}{3 \left( 1 - \left| U_{e3} \right|^2 \right)}
\ge \frac{1}{3},
\ee
which is somewhat disfavoured experimentally,
since the best-fit value for $\sin^2{\theta_\odot}$
is $0.304 < 1/3$~\cite{tortola}.
The situation becomes
worse
for the trimaximal-mixing hypothesis when $U_{e3}$ is nonzero;
indeed,
a recent fit~\cite{fogli} found $\left| U_{e3} \right|^2 = 0.016 \pm 0.010$,
in agreement with $\left| U_{e3} \right|^2 = 0.01 
\raisebox{2pt}{$\begin{array}{c} 
\scriptstyle +0.016 \\[-2.5mm] \scriptstyle -0.011 \end{array}$}$ 
in \cite{tortola},
which is not yet a significant indication for a nonzero $U_{e3}$.
In any case,
the trimaximal-mixing hypothesis
might be testable soon through more accurate measurements
of $\left| U_{e2} \right|$ and $\left| U_{e3} \right|$.

A trimaximal lepton mixing matrix $U$
has the moduli of two of its matrix elements \emph{of the same column} fixed.
This means that 
only two parameters remain in $U$,\footnote{Besides,
two Majorana phases are present in $U$
if the neutrinos are Majorana fermions.}
which can be taken as
$\left| U_{e3} \right|$,
or the mixing angle $\theta_{13}$ and a Dirac phase.
Clearly,
for the latter a convention has to be adopted.
Using the convention for the Dirac phase $\delta$ promulgated by~\cite{RPP},
we find
\be\label{tan}
\tan 2\theta_\mathrm{atm} = 
\frac{1 - 2 \left| U_{e3} \right|^2}{\left| U_{e3} \right| \cos{\delta} 
\sqrt{2 - 3 \left| U_{e3} \right|^2}}.
\ee

In the following,
we shall employ the following parameterization of a trimaximal mixing matrix:
%
%%%%
\be
U =
\mathrm{diag} \left(
e^{i \delta_e}, e^{i \delta_\mu}, e^{i \delta_\tau}
\right) \,
U_\mathrm{HPS} 
\left( \begin{array}{ccc}
c & 0 & s  e^{-i \psi} \\ 0 & 1 & 0 \\ -s e^{i \psi} & 0 & c
\end{array} \right)
\mathrm{diag} \left( e^{i \beta_1}, e^{i \beta_2}, e^{i \beta_3} \right),
\label{jftyr}
\ee
where $c = \cos{\theta}$ and $s = \sin{\theta}$.
The mixing angle $\theta$ parameterizes
how much lepton mixing deviates from tri-bimaximality.
The phase $\psi$ is of Dirac
type.\footnote{Note that the phase $\psi$ corresponds to
a Dirac phase convention different from that of $\delta$.}
The phases $\delta_{e, \mu, \tau}$,
together with one of the phases $\beta_j$,
are unphysical;
only the phase differences $2 \left( \beta_1 - \beta_2 \right)$
and $2 \left( \beta_2 - \beta_3 \right)$ can be physical,
if the neutrinos happen to be of Majorana type.
The modification~(\ref{jftyr}) of the HPS mixing matrix has recently
also been considered in~\cite{he}.

From equations~(\ref{jftyr}) and~(\ref{kgure}),
\ba
\left| U_{e3} \right|^2 &=& \frac{2}{3}\, s^2,
\label{jfuri}
\\
\sin^2{\theta_\mathrm{atm}}
&=& \frac{1}{2} + \frac{c s \cos{\psi}}{\sqrt{3}
\left( 1 - \left| U_{e3} \right|^2 \right)}.
\label{lprtq}
\ea
Therefore,
\be
\left( \sin^2{\theta_\mathrm{atm}} - \frac{1}{2} \right)^2
\le \frac{\left| U_{e3} \right|^2}{2}\,
\frac{1 - \frac{3}{2} \left| U_{e3} \right|^2}
{\left( 1 - \left| U_{e3} \right|^2 \right)^2}.
\label{udger}
\ee
This inequality can also be obtained directly from equation~(\ref{tan}).
The inequality~(\ref{udger}) relates,
when the mixing is trimaximal,
the maximal possible departure from maximal atmospheric-neutrino mixing,
\textit{i.e.}~from $\sin^2{\theta_\mathrm{atm}} = 1/2$,
to the value of $\left| U_{e3} \right|$.

\section{A simple model}
\label{model}

\paragraph{Introduction}
Let us assume that the neutrinos are Majorana fermions.
Then,
in the weak basis in which the charged-lepton mass matrix is diagonal,
the effective mass Lagrangian for the light neutrinos is
%
%%%%
\be
\mathcal{L}_\mathrm{neutrino \ mass} = \frac{1}{2}
\left( \begin{array}{ccc}
\nu_{eL}^T, & \nu_{\mu L}^T, & \nu_{\tau L}^T
\end{array} \right) C^{-1} \mnu
\left( \begin{array}{c}
\nu_{eL} \\ \nu_{\mu L} \\ \nu_{\tau L} \end{array} \right)
+ \mathrm{H.c.},
\ee
where $C$ is the Dirac--Pauli charge-conjugation matrix in Dirac space
and $\mnu$ is a $3 \times 3$ symmetric matrix in flavour space.
The lepton mixing matrix $U$ diagonalizes $\mnu$:
\be
U^T \mnu \, U = \mbox{diag} \left( m_1, m_2, m_3 \right),
\ee
the neutrino masses $m_{1,2,3}$ being non-negative real.
Using the parameterization of a trimaximal $U$ in equation~(\ref{jftyr}),
we shall denote
\be
\mu_j \equiv m_j e^{- 2 i \beta_j} \quad \mbox{for} \ j = 1, 2, 3.
\ee
Then,
if we assume the phases $\delta_{e,\mu,\tau}$ to vanish,
we have
\be
\mathrm{diag} \left( \mu_1, \mu_2, \mu_3 \right) =
\left( \begin{array}{ccc}
c & 0 & -s e^{i \psi} \\ 0 & 1 & 0 \\ s e^{-i \psi} & 0 & c
\end{array} \right)
\left( U_\mathrm{HPS}^T \mnu U_\mathrm{HPS} \right)
\left( \begin{array}{ccc}
c & 0 & s e^{-i \psi} \\ 0 & 1 & 0 \\ -s e^{i \psi} & 0 & c
\end{array} \right).
\label{udjcl}
\ee
Thus,
up to the phase transformation given by the phases $\delta_{e,\mu,\tau}$,
trimaximal mixing means that the vector $\left( 1, 1, 1 \right)^T$
is an eigenvector of $\mnu$ with eigenvalue $\mu_2$.
This means that,
in the phase convention $\delta_e = \delta_\mu = \delta_\tau = 0$
for $\mnu$,
the sum of the matrix elements of $\mnu$
%%%%
over all rows and columns of $\mnu$ is equal (to $\mu_2$).
It is the purpose of this section
to construct a model based on this idea.

\paragraph{The group $\Delta(27)$}
\label{delta27}

$\Delta(27)$ is a discrete group with 27 elements.
It has two inequivalent triplet irreducible representations (irreps),
$\ut$ and $\ut^\ast$,
and nine inequivalent singlet irreps,
$\uo^{(p,q)}$ ($p, q = 0, 1, 2$).
The triplet irreps of $\Delta(27)$ are faithful,
the singlet irreps are non-faithful.
The group $\Delta(27)$ is generated by two transformations,
$C$ and $T$.
In the $\ut$,
those transformations are represented as
\be
\ut: \quad 
C \to \left( \begin{array}{ccc}
0 & 0 & 1 \\ 1 & 0 & 0 \\ 0 & 1 & 0
\end{array} \right),
\quad
T \to \left( \begin{array}{ccc}
1 & 0 & 0 \\ 0 & \omega & 0 \\ 0 & 0 & \omega^2
\end{array} \right),
\quad \mbox{with} \ \omega \equiv e^{2\pi i/3} = 
\frac{-1 + i \sqrt{3}}{2}.
\label{kjdxz}
\ee
Notice that
the matrices representing $C$ and $T$ belong to $SU(3)$,
therefore $\Delta(27)$ may be viewed as a subgroup of $SU(3)$.
In the $\ut^\ast$,
\be
\ut^\ast: \quad 
C \to \left( \begin{array}{ccc}
0 & 0 & 1 \\ 1 & 0 & 0 \\ 0 & 1 & 0
\end{array} \right),
\quad
T \to \left( \begin{array}{ccc}
1 & 0 & 0 \\ 0 & \omega^2 & 0 \\ 0 & 0 & \omega
\end{array} \right).
\ee
In the singlet irreps,
\be
\uo^{(p,q)}: \quad C \to \omega^p, \quad T \to \omega^q.
\ee
The relevant tensor products of irreps of $\Delta(27)$ are
\ba
\label{tensor1}
\ut \otimes \ut &=& \ut^\ast \oplus \ut^\ast \oplus \ut^\ast,
\\
\label{tensor2}
\ut \otimes \ut^\ast &=& \oplus^2_{p,q=0} \ \uo^{(p,q)}.
\ea

\paragraph{Multiplets and symmetries}

In our model we consider only the lepton sector
and the electroweak interactions.
The gauge group is the standard $SU(2) \times U(1)$.
There are three left-handed-lepton doublets
$D_{\alpha L} = \left( \nu_{\alpha L}, \, \alpha_L \right)^T$
and three right-handed charged-lepton singlets $\alpha_R$
($\alpha = e, \mu, \tau$).
We add to these standard multiplets four right-handed neutrino singlets
in order to enable the seesaw mechanism~\cite{seesaw}.
Those four right-handed neutrinos are divided in two sets,
three $\nu_{\alpha R}$ and one $\nu_{0R}$.
In the scalar sector,
there are four Higgs doublets,
once again divided in two sets:
three $\phi_\alpha$ and one $\phi_0$.
We need moreover a scalar gauge singlet $S$.
The $SU(2) \times U(1)$ and $\Delta(27)$ assignments
of all these multiplets are given in table~\ref{table}. 
\begin{table}
\begin{center}
\begin{tabular}{c|cccc|ccc}
& $D_{\alpha L}$ & $\alpha_R$ & $\nu_{\alpha R}$ & $\nu_{0R}$ &
$\phi_\alpha$ & $\phi_0$ & $S$
\\ \hline
$SU(2) \times U(1)$ & $\left( \utwo, -1 \right)$ &
$\left( \uo, -2 \right)$ & $\left( \uo, 0 \right)$ &
$\left( \uo, 0 \right)$ & $\left( \utwo, 1 \right)$ &
$\left( \utwo, 1 \right)$ & $\left( \uo, 0 \right)$ \\
$\Delta(27)$ & $\ut$ & $\ut^\ast$ & $\ut$ & $\uo^{(1,0)}$ &
$\ut^\ast$ & $\uo^{(0,0)}$ & $\uo^{(1,0)}$ 
\end{tabular}
\end{center}
\caption{Fermion and scalar multiplets of our model
\label{table}}
\end{table}

\paragraph{Additional $\zz$ symmetries}
Besides the gauge group and $\Delta(27)$,
we impose three extra $\zz$ symmetries $\mathbf{z}_{e, \mu, \tau}$:
\be
\label{z}
\mathbf{z}_\alpha: \quad 
\alpha_R \to - \alpha_R, \
\phi_\alpha \to - \phi_\alpha,
\ee
while all other multiplets remain unchanged.
Each $\mathbf{z}_\alpha$ has the purpose of
``gluing''
$\alpha_R$ to $\phi_\alpha$ in the Yukawa couplings;
this is the same idea as in~\cite{GL06} 
(see also~\cite{nasri}). 
Since the $\mathbf{z}_\alpha$ do not commute with $\Delta(27)$,
the horizontal symmetry group employed in our model
is actually much larger than $\Delta(27)$.

\paragraph{Yukawa couplings}

Let us first consider the Yukawa couplings of the $\alpha_R$.
They are
\be
\mathcal{L}_{\alpha_R \, \mathrm{Yukawas}} =
- y_1 \sum_{\alpha = e, \mu, \tau}
\bar D_{\alpha L} \alpha_R \phi_\alpha
+ \mathrm{H.c.}
\label{hftry}
\ee
According to equation~(\ref{tensor1}),
$\Delta(27)$ would allow two other couplings,
\be
\begin{array}{l}
- y_2 \left(
\bar D_{eL} \mu_R \phi_\tau
+ \bar D_{\mu L} \tau_R \phi_e
+ \bar D_{\tau L} e_R \phi_\mu
\right)
\\
- y_3 \left(
\bar D_{eL} \tau_R \phi_\mu
+ \bar D_{\mu L} e_R \phi_\tau
+ \bar D_{\tau L} \mu_R \phi_e
\right) + \mathrm{H.c.}
\end{array}
\label{urtsy}
\ee
These terms would destroy trimaximal mixing.
It is for this reason that
we have introduced into our model the symmetries $\mathbf{z}_\alpha$,
which remove the terms~(\ref{urtsy}) from the Lagrangian.
The masses of the charged leptons
are $m_\alpha = \left| y_1 v_\alpha \right|$,
where $v_\alpha$ is the VEV of the neutral component of $\phi_\alpha$.
If we manage $v_{e, \mu, \tau}$ to be all different,
then the charged leptons will be non-degenerate in mass as desired.

The Yukawa couplings of the right-handed neutrinos are given by
\be
\label{nuRcoupling}
\mathcal{L}_{\nu_R \, \mathrm{Yukawas}} =
- y_4 \sum_{\alpha = e, \mu, \tau}
\bar D_{\alpha L} \nu_{ \alpha R}
\left( i \tau_2 \phi_0^\ast \right)
+ \frac{y_5}{2}\, \nu_{0R}^T C^{-1} \nu_{0R} \, S
+ \mathrm{H.c.}
\ee

\paragraph{Soft breaking of the symmetries}

Soft breaking of (super)symmetries
plays an important role in many models.
Soft breaking is usually needed in models which
want to explain mixing features through some symmetries.
It has been emphasized that
successful models need a residual symmetry~\cite{blum};
for instance,
in~\cite{GL06,GL01} the residual symmetry after soft breaking
is the $\mu$--$\tau$ interchange symmetry,
which leads to maximal atmospheric-neutrino mixing
and to $U_{e3} = 0$.

In the present model,
we break $\Delta(27)$ softly in two steps.
Firstly we allow it to be broken,
by terms of dimension three,
down to the $\zzz$ symmetry generated by the transformation $C$,
which is denoted $\zzz (C)$.
Secondly we allow $\zzz (C)$ to be softly broken
by terms of dimension two.
We thus have the soft-breaking chain
\be
\label{soft}
\Delta(27)\, \stackrel{\dim 3}{\longrightarrow}\, \zzz(C)\, 
\stackrel{\dim 2}{\longrightarrow}\, \{ e \},
\ee
where $\{ e \}$ symbolizes the trivial group
consisting only of the unit element.

The soft-breaking terms of dimension three
occur in the Lagrangian of bare Majorana masses
\ba
\mathcal{L}_\mathrm{Majorana \ masses} &=&
\frac{M_0^\ast}{2} \sum_{\alpha = e, \mu, \tau}
\nu_{\alpha R}^T C^{-1} \nu_{\alpha R}
\no & &
+ M_1^\ast \left(
\nu_{eR}^T C^{-1} \nu_{\mu R}
+ \nu_{\mu R}^T C^{-1} \nu_{\tau R}
+ \nu_{\tau R}^T C^{-1} \nu_{eR}
\right)
\no & &
+ \frac{M_2^\ast}{2} \left(
\nu_{eR}^T + \omega \, \nu_{\mu R}^T + \omega^2 \, \nu_{\tau R}^T
\right) C^{-1} \nu_{0R} + \mathrm{H.c.}
\label{LM}
\ea
The soft-breaking terms of dimension two occur in the scalar potential
\be
V =
c_e \phi_e^\dagger \phi_e
+ c_\mu \phi_\mu^\dagger \phi_\mu
+ c_\tau \phi_\tau^\dagger \phi_\tau
+ \cdots,
\ee
the coefficients (with dimension mass squared) $c_e$,
$c_\mu$ and $c_\tau$ being all different,
thereby breaking $\zzz (C)$.
This is needed in order to obtain,
upon spontaneous symmetry breaking,
different VEVs $v_{e,\mu,\tau}$ and,
therefore,
different charged-lepton masses:
\be
m_e : m_\mu : m_\tau
= \left| v_e \right| : \left| v_\mu \right| : \left| v_\tau \right|.
\ee
Furthermore, 
we might include in $V$ all terms like $\phi_e^\dagger \phi_\mu$,
etc.
This would also break the symmetries $\mathbf{z}_\alpha$ softly
and would avoid all potential problems
with spontaneous breaking of discrete symmetries in our model.

\paragraph{Seesaw mechanism}

From equations~(\ref{nuRcoupling}) and~(\ref{LM}),
we see that there are in our model
Majorana and Dirac neutrino mass matrices
\be
\label{MRD}
M_R = \left( \begin{array}{cccc}
M_0 & M_1 & M_1 & M_2 \\
M_1 & M_0 & M_1 & \omega^2\, M_2 \\
M_1 & M_1 & M_0 & \omega\, M_2 \\
M_2 & \omega^2\, M_2 & \omega\, M_2 & M_N
\end{array} \right),
\quad
M_D = \left( \begin{array}{ccc}
a & 0 & 0 \\ 0 & a & 0 \\ 0 & 0 & a \\ 0 & 0 & 0 
\end{array} \right),
\ee
respectively.
We have defined 
$M_N = y_5^\ast v_S^\ast$,
where $v_S$ is the VEV of the scalar singlet $S$,
and $a = y_4^\ast v_0$,
where $v_0$ is the VEV of the neutral component of $\phi_0$.
The seesaw mechanism~\cite{seesaw}
tells us that
\be
\label{see}
\mnu = - M_D^T M_R^{-1} M_D.
\ee
After some algebra
one finds that $\mnu$ is of the form
\be
\label{mnu}
\mnu = 
\left( \begin{array}{ccc}
x + y & z + \omega^2 y &  z + \omega y \\
z + \omega^2 y & x + \omega y & z + y \\
z + \omega y & z + y & x + \omega^2 y
\end{array} \right),
\ee
where
\ba
x &=& - a^2\, \frac{M_0 + M_1}
{\left( M_0 - M_1 \right) \left( M_0 + 2 M_1 \right)}\, ,
\label{x} \\
z &=& a^2\, \frac{M_1}
{\left( M_0 - M_1 \right) \left( M_0 + 2 M_1 \right)}\, ,
\label{zzzz} \\
y &=& - a^2\, \frac{M_2^2}{M_N \left( M_0 - M_1 \right)^2}\, .
\label{y}
\ea
It is clear that
\be
\label{eigen}
\mnu \left( \begin{array}{c} 1 \\ 1 \\ 1 \end{array} \right)
= \left( x + 2 z \right)
\left( \begin{array}{c} 1 \\ 1 \\ 1 \end{array}
\right).
\ee
Therefore,
our model predicts trimaximal mixing.

\paragraph{A further prediction}

We compute
\be
\label{partial}
U_\mathrm{HPS}^T \mnu U_\mathrm{HPS} = 
\left( \begin{array}{ccc}
x - z + \frac{3}{2}\, y & 0 & i\, \frac{3}{2}\, y \\
0 & x + 2 z & 0 \\
i\, \frac{3}{2}\, y & 0 & x - z - \frac{3}{2}\, y 
\end{array} \right).
\ee
Comparing this result with equation~(\ref{udjcl}),
we see that
\be
\mu_2 = x + 2 z = \frac{- a^2}{M_0 + 2 M_1}
\ee
and 
\ba
x - z + \frac{3}{2}\, y
&=& \mu_1 c^2 + \mu_3 s^2 e^{2 i \psi},
\\
x - z - \frac{3}{2}\, y
&=& \mu_3 c^2 + \mu_1 s^2 e^{- 2 i \psi},
\\
i\, \frac{3}{2}\, y
&=& c s \left( \mu_3 e^{i \psi} - \mu_1 e^{- i \psi} \right).
\ea
Therefore,
\be
\frac{\mu_1}{\mu_3} =
\left( \frac{c - i s e^{i \psi}}{c - i s e^{- i \psi}} \right)^2,
\ee
hence
\be
\frac{m_1}{m_3} = \frac{1 + 2 c s \sin{\psi}}{1 - 2 c s \sin{\psi}}.
\ee
Comparing this result with equations~(\ref{jfuri}) and~(\ref{lprtq}),
one finds that
\be
\left| U_{e3} \right|^2 \left( 2 - 3 \left| U_{e3} \right|^2 \right)
- \left( 1 - \left| U_{e3} \right|^2 \right)^2
\left( 2\, \sin^2{\theta_\mathrm{atm}} - 1 \right)^2
= \frac{1}{3}\, \frac
{\left( m_3^2 - m_1^2 \right)^2}
{\left( m_1 + m_3 \right)^4},
\label{njdfw}
\ee
\textit{cf.}~inequality~(\ref{udger}).
The prediction~(\ref{njdfw}) relates
the deviation from tri-bimaximal lepton mixing
to the mass ratio $m_1/m_3$. 
Using the experimental value of $\datm$,
then,
the sum of neutrino masses $m_1 + m_3$
is determined by the deviation from tri-bimaximal mixing.

With the experimental $3 \, \sigma$ bounds~(\ref{urtye}) and~(\ref{ndunf}),
one finds
\be\label{lb}
m_1 + m_3 \ge 0.060\, \mathrm{eV}.
\ee
If $\left| U_{e3} \right|^2$ is smaller than the bound~(\ref{urtye})
and/or if $\sin^2{\theta_\mathrm{atm}}$ deviates from $1/2$,
%%%%
then the lower bound~(\ref{lb}) on $m_1 + m_3$ is strengthened.
However,
the fourth power of $m_1 + m_3$ in equation~(\ref{njdfw}) 
dampens this effect.

\begin{figure}[t]
\begin{center}
\epsfig{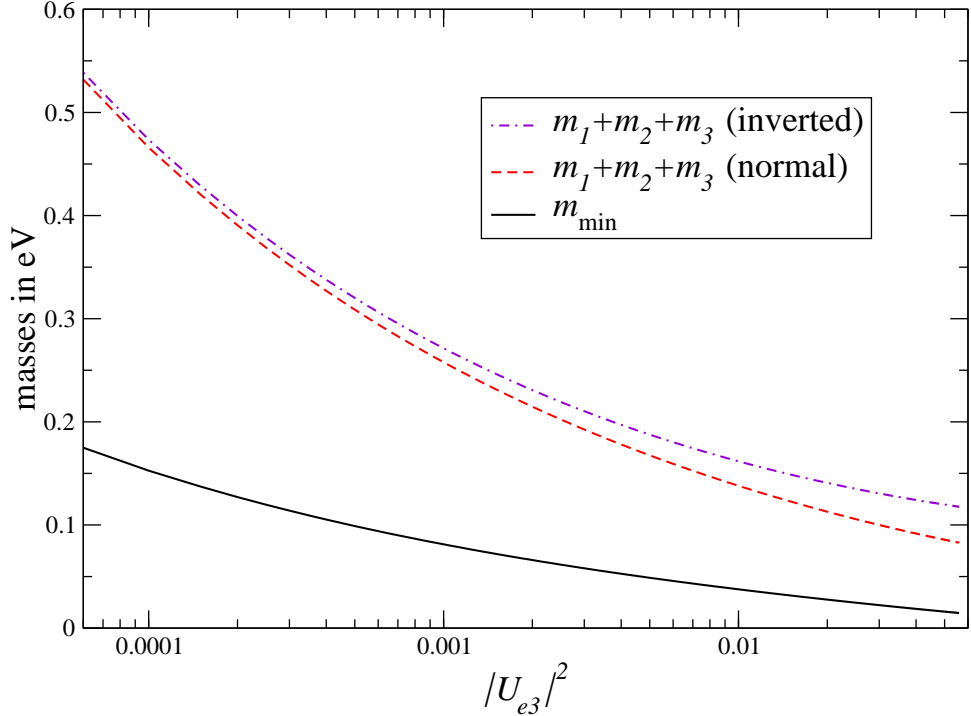}
\end{center}
\caption{The minimal neutrino mass and the sum of the neutrino
  masses, for both types of spectra, as a function of 
  $\left| U_{e3} \right|^2$. We have fixed the mass-squared
  differences at the mean values given in~\cite{tortola} and
  assumed that atmospheric mixing is maximal.
\label{mfig}}
\end{figure}
Taking the experimental values of $\datm$ and $\dsol$ as input,
equation~(\ref{njdfw}) determines the smallest neutrino mass
$m_\mathrm{min}$, which is $m_1$ for
the normal and $m_3$ for the inverted spectrum, as a function of 
$\left| U_{e3} \right|^2$ and $\sin^2{\theta_\mathrm{atm}}$.
In Fig.~\ref{mfig} we have depicted $m_\mathrm{min}$ as a function of 
$\left| U_{e3} \right|^2$, fixing $\sin^2{\theta_\mathrm{atm}}$ at 0.5
and using the mean values
$\datm = 2.4 \times 10^{-3}$ eV$^2$ and 
$\dsol = 7.65 \times 10^{-5}$ eV$^2$
from~\cite{tortola}. 
We also show the sum of the neutrino
%%%%
masses for both the normal and the inverted spectra.
We see that at large $\left| U_{e3} \right|^2$ the sum of the neutrino
masses is safely below present cosmological bounds~\cite{raffelt}.

\paragraph{Parameter counting and the number of predictions}

The neutrino mass matrix~(\ref{mnu}) is a five-parameter mass matrix
because it has three complex parameters, with only their relative
phases having a physical meaning. One can easily show with
equation~(\ref{partial}) that $\mnu^\dagger \mnu$
has only four parameters. The neutrino masses, the mixing angles and
the Dirac phase follow all from $\mnu^\dagger \mnu$, only for the
investigation of the Majorana phases we need in fact $\mnu$. 

Therefore, the four parameters in $\mnu^\dagger \mnu$ determine
seven observables. As a consequence, there must be three
predictions. Indeed, those predictions are given by
equations~(\ref{solmix}) and (\ref{tan}), which follow from trimaximal
mixing alone, and equation~(\ref{njdfw}), which is a result of our
specific model.

As for $\mnu$, the difference of the two Majorana phases can be
expressed in terms of its five parameters; this constitutes the
additional prediction if we consider $\mnu$ instead of 
$\mnu^\dagger \mnu$. However, in our model, we expect no significant
result for the effective neutrino mass in neutrinoless double
$\beta$-decay as compared to the general case, since one Majorana
phase is competely free and $\left| U_{e3} \right|^2$ is small.

\section{Variations on the model}
\label{extensions}

\subsection{Use of a $CP$ symmetry}

The model presented in the previous section
does not possess $\mu$--$\tau$ interchange symmetry,
which would require,
in the $\mnu$ of equation~(\ref{mnu}),
$y = 0$,
leading to two degenerate neutrinos.
Alternatively,
though,
we may impose on the model a $CP$ symmetry which interchanges
the $\mu$ and $\tau$ families~\cite{CPpaper},
\textit{viz.}
\be\label{CP}
\left( \begin{array}{c}
\nu_{e R} \\ \nu_{\mu R} \\ \nu_{\tau R}
\end{array} \right) \left( t, \vec r\, \right)
\stackrel{CP}{\longrightarrow}
i\,\gamma_0\, C
\left( \begin{array}{c}
\bar \nu_{e R}^T \\ \bar \nu_{\tau R}^T \\ \bar \nu_{\mu R}^T
\end{array} \right) \left( t, - \vec r\, \right),
\quad
\nu_{0 R} \left( t, \vec r\, \right)
\stackrel{CP}{\longrightarrow}
i\,\gamma_0\, C \,
\bar \nu_{0R}^T \left( t, - \vec r\, \right),
\ee
and so on.
Such a $CP$ symmetry would force $M_{0,1,2}$ 
in equation~(\ref{LM}) to be real,
hence $x$,
$y$ and $z$ in equations~(\ref{mnu})--(\ref{y}) 
to be real.
One would thus obtain a neutrino mass matrix with three parameters only,
which fulfils
\be\label{CPrelation}
S \mnu S = \mnu^*
\quad \mbox{with} \quad
S = \left( \begin{array}{ccc}
1 & 0 & 0 \\ 0 & 0 & 1 \\ 0 & 1 & 0
	   \end{array} \right).
\ee
As shown in~\cite{CPpaper}, a neutrino mass matrix obeying 
equation~(\ref{CPrelation}) predicts $\sin^2{\theta_\mathrm{atm}} = 1/2$,
maximal $CP$ violation,
\textit{i.e.}~$e^{i\delta} = \pm i$, 
and vanishing Majorana phases.
Therefore,
a restricted version of our model,
including the $CP$ symmetry~(\ref{CP}),
has these predictions in addition to those of trimaximal mixing.

\subsection{One more right-handed neutrino}

If one wants to have trimaximal mixing
without the extra prediction~(\ref{njdfw}),
one may introduce into the model one more right-handed neutrino,
$\nu_{0R}^\prime$,
transforming as $\uo^{(2,0)}$ 
under $\Delta(27)$.
This leads to one extra term
\be
\frac{y_6}{2}\, \nu_{0R}^{\prime T} C^{-1} \nu_{0R}^\prime \, S^\ast
+ \mathrm{H.c.}
\ee
%
%%%%% 
on
the right-hand side of equation~(\ref{nuRcoupling}),
and to two extra couplings
\be
\frac{M_3^\ast}{2} \left(
\nu_{eR}^T + \omega^2 \, \nu_{\mu R}^T + \omega \, \nu_{\tau R}^T
\right) C^{-1} \nu_{0R}^\prime
+ M_4^\ast \nu_{0R}^T C^{-1} \nu_{0R}^\prime
+ \mathrm{H.c.}
\ee
%
%%%%  
on
the right-hand side of equation~(\ref{LM}).
Equations~(\ref{MRD}) would then read
\be
M_R = \left( \begin{array}{ccccc}
M_0 & M_1 & M_1 & M_2 & M_3 \\
M_1 & M_0 & M_1 & \omega^2\, M_2 & \omega\, M_3 \\
M_1 & M_1 & M_0 & \omega\, M_2 & \omega^2\, M_3 \\
M_2 & \omega^2\, M_2 & \omega\, M_2 & M_N & M_4 \\
M_3 & \omega\, M_3 & \omega^2\, M_3 & M_4 & M_N^\prime
\end{array} \right),
\quad
M_D = \left( \begin{array}{ccc}
a & 0 & 0 \\ 0 & a & 0 \\ 0 & 0 & a \\ 0 & 0 & 0 \\ 0 & 0 & 0 
\end{array} \right),
\ee
with $m_N^\prime = y_6^\ast v_S$.
Instead of equation~(\ref{mnu}) one would then have
\be
\mnu = 
\left( \begin{array}{ccc}
x + y + t & z + \omega^2 y + \omega t &  z + \omega y + \omega^2 t \\
z + \omega^2 y + \omega t & x + \omega y + \omega^2 t & z + y + t \\
z + \omega y + \omega^2 t & z + y + t & x + \omega^2 y + \omega t
\end{array} \right).
\ee
This still predicts trimaximal mixing
but the extra prediction~(\ref{njdfw}) disappears.

\subsection{Use of $\Delta(3n^2)$ or other symmetry groups}

We may generalize the symmetry $T$ by using,
instead of equation~(\ref{kjdxz}),
\be
\ut: \quad
C = \left( \begin{array}{ccc}
0 & 0 & 1 \\ 1 & 0 & 0 \\ 0 & 1 & 0
\end{array} \right),
\quad
T = \left( \begin{array}{ccc}
1 & 0 & 0 \\ 0 & \sigma & 0 \\ 0 & 0 & \sigma^\ast
\end{array} \right),
\quad \mbox{with} \ \sigma = e^{2 \pi i / n} 
\quad (n \geq 3).
\ee
The transformation properties of the multiplets
%%%%
under $T$ would then be as shown in table~\ref{general};
fields not shown in that table transform trivially under $T$.
\begin{table}
\begin{center}
\begin{tabular}{c|ccc|c}
& $D_{L \alpha}$ & $\alpha_R$ & $\nu_{\alpha R}$ & $\phi_\alpha$ \\ \hline
$T$ & $T$ & $T^*$ & $T$ & $T^2$
\end{tabular}
\end{center}
\caption{Generalizing $T$. \label{general}}
\end{table}
All fields transform under $C$
in the same way as in section~\ref{model}. 
Apart from $T$,
all other details of the model are the same,
in particular the soft breaking as given by equation~(\ref{soft}).
Thus,
the matrix $M_R$ would remain unchanged,
since its form is fixed by the transformation $C$.

Let us consider $n \geq 4$.
Then the main conclusions are the following:
\begin{itemize}
\item The fermionic sector and,
therefore, 
the matrix $\mnu$ is the same for all $n$,
as given in equation~(\ref{MRD}).
\item The terms in equation~(\ref{urtsy}) are automatically forbidden,
therefore the $\zz$ symmetries of equation~(\ref{z}) are not needed.
\item The symmetry group is $\Delta (3n^2)$,
softly broken by terms of dimension three to $\zzz(C)$.
\end{itemize}
A detailed discussion of $\Delta (3n^2)$ is given in~\cite{luhn}.
Actually,
the terms of dimension four in the Lagrangian
are invariant under \emph{all} the permutations.
This leads to the symmetry group $\Delta (6n^2)$---see~\cite{luhn}. 

There is still another way to produce the present model.
Consider cyclic permutations (or all permutations),
plus family lepton-number symmetries $U(1)_\alpha$
and the $\zz$ symmetries of equation~(\ref{z}).
The scalar doublets carry no lepton number.
The neutrino $\nu_{0R}$ and the scalar singlet $S$
may either carry lepton number or not.
The $U(1)_\alpha$ are softly broken by terms of dimension three,
the residual symmetry being once again $\zzz(C)$. 

All the groups considered here produce identical models
as far as the terms in the Lagrangian involving the fermion fields
are concerned;
the only differences which may arise
occur in the terms of dimension four in the scalar potential.

\section{Conclusions}
\label{conclusions}

In this paper we have focused our attention on trimaximal lepton mixing,
with a two-parameter lepton mixing matrix, 
which generalizes tri-bimaximal mixing. 
Our main motivation was to allow for a deviation of 
$\left| U_{e3} \right|^2$ from zero; recent studies point out that
possibility~\cite{fogli,tortola}. 
Trimaximal mixing correlates the deviation of $\left| U_{e3} \right|^2$
from zero with a deviation of $\sin^2{\theta_\odot}$ from 1/3 and of
$\sin^2{\theta_\mathrm{atm}}$ from 1/2---see equations~(\ref{solmix})
and (\ref{tan}).
A particular consequence is $\sin^2{\theta_\odot} \geq 1/3$, which is slightly
disfavoured by the data at the moment, but in any case might be tested
soon. 

We have also constructed a seesaw model (or rather a class of models) where
trimaximal mixing is enforced by a family symmetry group. In this
model, the mass matrix of the light neutrinos, given by
equation~(\ref{mnu}), has five physical parameters; it includes not
only the predictions of trimaximal mixing but also
equation~(\ref{njdfw}) which relates the deviation from tri-bimaximal
lepton mixing to the ratio $m_1/m_3$ of neutrino masses. As for a family
symmetry group, we have considered several possibilities; one of the
most straightforward ones is based on the group $\Delta(27)$.
We have also considered a restricted version of our model by
imposing, in addition, a non-standard $CP$ transformation; in this way
we are lead to a three-parameter neutrino mass matrix which predicts
not only trimaximal mixing but also
$\sin^2 \theta_\mathrm{atm} = 1/2$.

Our model has some peculiarities, like the need of four (or more)
right-handed neutrino singlets; the fourth neutrino singlet couples to the 
the three usual ones denoted by $\nu_{\alpha R}$ ($\alpha =
e,\,\mu,\,\tau$) via a scalar gauge singlet. In the mass terms of the 
$\nu_{\alpha R}$, the symmetry group is broken softly down to a $\zzz$. 
An outstanding feature of the model is that no vacuum alignment is
required, despite its scalar content of four Higgs doublets and the
scalar singlet. This is to be contrasted with models for tri-bimaximal
mixing which are plagued by the intricacies of vacuum alignment.

\paragraph{Acknowledgements} 
W.G.~thanks F.~Bazzocchi and S.~Morisi for stimulating discussions.
The work of L.L.~was supported by the Portuguese
\textit{Funda\c c\~ao para a Ci\^encia e a Tecnologia}
through the project U777--Plurianual.
W.G.~and~L.L.~acknowledge support from EU under the
MRTN-CT-2006-035505 network programme.

\end{document}